\documentclass[aps,english,prb,reprint,superscriptaddress]{revtex4-2}

\usepackage{graphicx}
\usepackage{float}
\usepackage{amssymb}
\usepackage{babel}
\usepackage{color}
\usepackage{amsmath}
\usepackage{amsfonts}

\begin{document}

\title{Simple linear response model for predicting energy band alignment of two-dimensional vertical heterostructure}

\author{Javad G. Azadani}\thanks{These authors contributed equally to this work.}
\affiliation{Department of Electrical and Computer Engineering, University of Minnesota, Minneapolis, Minnesota 55455, USA}

\author{Seungjun Lee}\thanks{These authors contributed equally to this work.}
\affiliation{Department of Physics, Kyung Hee University, Seoul 02447, Korea}

\author{Hyeong-Ryul Kim}
\affiliation{Department of Physics, Kyung Hee University, Seoul 02447, Korea}

\author{Hussain Alsalman}
\affiliation{Department of Electrical and Computer Engineering, University of Minnesota, Minneapolis, Minnesota 55455, USA}
\affiliation{King Abdulaziz City for Science and Technology (KACST), Riyadh 6086-11442, Kingdom of Saudi Arabia}

\author{Young-Kyun Kwon}\email{ykkwon@khu.ac.kr}
\affiliation{Department of Physics, Kyung Hee University, Seoul 02447, Korea}

\author{Jerry Tersoff}
\affiliation{IBM T. J. Watson Research Center, Yorktown Heights, New York 10598, USA}

\author{Tony Low}\email{tlow@umn.edu}
\affiliation{Department of Electrical and Computer Engineering, University of Minnesota, Minneapolis, Minnesota 55455, USA}
\affiliation{Department of Physics, Kyung Hee University, Seoul 02447, Korea}

\begin{abstract}
The Anderson and midgap models are often used in the study of semiconductor heterojunctions, but for van der Waals (vdW) vertical heterostructures they have shown only very limited success. Using the group-IV monochalcogenide vertical heterostructures as a prototypical system, we propose a linear response model and compare the effectiveness of these models in predicting density functional theory (DFT) band alignments, band types and bandgaps. We show that the DFT band alignment is best predicted by the linear response model, which falls in between the Anderson and midgap models. Our proposed model can be characterized by an interface dipole $\alpha\times(E_{m2}-E_{m1})$, where the linear response coefficient $\alpha$ = 0 and 1 corresponds to the Anderson and midgap model respectively, and $E_{m}$ is the midgap energy of the monolayer, which can be viewed as an effective electronegativity. For group-IV monochalcogenides, we show that $\alpha$ = 0.34 best captures the DFT band alignment of the vdW heterostructure, and we discuss the viability of the linear response model considering other effects such as strains and band hybridization, and conclude with an application of the model to predict experimental band alignments. 
\end{abstract}

\maketitle

\section{Introduction}
 Energy band alignment in semiconductor heterostructures is one of the most important properties in designing electronic and optoelectronic devices \cite{tsai2014monolayer,cheng2014electroluminescence,hong2014ultrafast,roy20162d}. A great deal of work has been directed towards the understanding of band-discontinuity at semiconductor heterojunction for several decades since the 1970s \cite{anderson1988experiments,anderson1960germanium,harrison1988elementary,tersoff1984theory,tersoff1986band,van1989band,van1987theoretical,frensley1977theory,tejedor1978simple,langer1985deep,alferov1998history}. These works span a wide spectrum, with simple and physically motivated models that predicts bands lineup of heterostructure based on the electronic properties of its constituent bulk semiconductors \cite{anderson1988experiments,tersoff1984theory,tejedor1978simple,langer1985deep}, while others approached it using elaborate first principle self-consistent atomistic calculations based on density function theory (DFT), or semi-empirical pseudopotential, or local atomic orbitals methods \cite{harrison1988elementary,van1989band,van1987theoretical,frensley1977theory}. These approaches have facilitated the understanding and design of III-V and II-VI semiconductor heterojunctions and played a crucial role in providing insights to experimental measurements and devices.  
 
 Heterojunction band alignment refers to the relative band edge energies of the respective semiconductors at the interface. The most elementary model was provided by Anderson \cite{anderson1988experiments,anderson1960germanium}, which states that the vacuum levels of the two semiconductors should be aligned on both sides of the junction. Hence, the conduction band offset is given by the difference in electron affinities of the two materials and the valence band offset is the sum of conduction band offset and bandgap difference. However, this model may not be well obeyed in practice, due to formation of a charge dipole at the interface. Studies of conventional bulk semiconductor heterostructures have attributed the physical origin of this interface dipole to various factors, which includes interface states \cite{rehr1974wannier,goetzberger1976interface}, gap states induced by wave function penetration \cite{tersoff1984theory}, band hybridization \cite{koda2018trends}, among other effects. Compared to conventional semiconductors, band alignments in atomically thin 2D van der Waals (vdW) heterostructures remain poorly understood.

 During the past few years, several studies have been conducted to study and predict the band alignments in vdW heterostructures \cite{pierucci2016band,latini2017interlayer,zhang2016systematic,you2016black,pontes2018layer,gong2013band,bellus2017type,cui2019electronic,li2017wse2,wei2015electronic,kim2015band,chiu2015determination,kang2013band,pham2019vertical,ozcelik2016band,wilson2017determination,chiu2017band,koda2017tuning,koda2018trends}, particularly heterostructures made by mechanical transfer \cite{akinwande2014two}. These studies found limited success of the Anderson model when compared to DFT results and experimental measurements \cite{chiu2017band}. The discrepancies have been ascribed to the band hybridization in heterostructure \cite{koda2017tuning}, and quantum dipoles \cite{koda2018trends}, and depends also on layer thickness \cite{leenaerts2016system}. 
 
In this work, we propose a linear response model. If the interlayer coupling is sufficiently weak, the heterostructure is characterized simply by the dipole formed. As in other models, our prediction is based on the properties of the separate constituent monolayers. However, it contains one additional parameter to describe the linear response. We focus on the family of group-IV monochalcogenides heterostructures as a prototypical system, which was shown to form mechanically stable and rigid heterostructures \cite{ozccelik2018tin}. Few-layer group-IV monochalcogenides have been synthesized \cite{antunez2011tin,ma2014growth,li2013single}, and their monolayers have been shown to be stable and can be exfoliated from their bulk phase \cite{patel2018growth}. Because of the reduced symmetry, they show piezoelectricity with large ionic dielectric screening and piezotronics for energy harvesting \cite{gomes2015enhanced,fei2015giant}, anisotropic thermal and electrical conductivity \cite{zhao2014ultralow,shafique2017thermoelectric} with high thermoelectric efficiency \cite{wang2015thermoelectric,guo2017thermoelectric}, and topological electronic properties \cite{liu2014spin,tanaka2012experimental,liu2015crystal}. The understanding of band alignments in these heterostructures would open new opportunities for designing novel materials optimized for particular applications. 

The organization of the paper is as follows. In Sec. II, we introduce our proposed linear response model. 
In Sec. III, we describe the atomic structure of group-IV monochalcogenides monolayers and its heterostructures, the computational method, and the estimation of a linear response coefficient for these materials. In Sec. IV, we review the Anderson and midgap band models, and compare them against the proposed linear response model using DFT results as a reference. In Sec. V, we survey the heterostructures of the group-IV monochalcogenides family and demonstrate the higher efficacy of our proposed linear response model in predicting bandgaps and band alignment types of 2D heterostructures, compared to the Anderson and midgap models. In Sec. VI, as an application, we apply the linear response model to predict band alignments of unstrained heterostructures, such as those formed by mechanical transfer techniques \cite{akinwande2014two}. Finally, Sec. VII sums up the conclusions of our study.

\section{Linear response model} 
If we could bring together two monolayers without letting them interact, and without allowing any charge transfer, then the vacuum energy could serve as a reference: the conduction band offset would be given by the difference in electron affinity of the respective monolayers \cite{anderson1988experiments,anderson1960germanium}. However, in general the layers do interact, and there is some charge rearrangement, leading to an overall dipole moment of the bilayer. An accurate calculation of this dipole requires a full density function calculation, which is not feasible for incommensurate structures. We therefore consider a widely used approximation, that each separate layer is characterized by a “neutrality level” (here take as the midgap), such that the distance from the vacuum level to the neutrality level is a measure of the electronegativity \cite{mulliken1934new,mulliken1935electronic,tersoff1985schottky}, and charge moves toward the more electronegative layer. We also make a second approximation of linearity, that the charge transfer and resulting dipole are linear in the neutrality-level difference:  
\begin{equation}
eV_{h} = \beta(E'_{m2}-E'_{m1})
\end{equation}
where $E’_{m1}$ and $E’_{m2}$ are the midgap energies of the constituent monolayers, and $\beta$ is a dimensionless parameter to be determined. 
If we write the midgap energy of the separate monolayers, relative to vacuum, as $E_{m1}$ and $E_{m2}$, then their offset is given by the non-interacting (``Anderson model’’) offset plus this dipole:    
\begin{equation}
E’_{m2}-E’_{m1} = (E_{m2}-E_{m1})-eV_{h}
\end{equation}

Solving Eqs. (1) and (2) self-consistently gives the linear response model as
\begin{equation}
eV_{h} = \alpha (E_{m2}-E_{m1})
\end{equation}
where $\alpha$ = $\beta$/($\beta$+1).

We note that the limiting case of $\alpha=0$ and 1 corresponds to the well-known Anderson and midgap band alignment models, respectively.

Neglecting charge transfer and dipole formation correspond to $\beta = 0$, so $\alpha= 0$, and the vacuum levels of two semiconductors are aligned on either side of the junction. This corresponds to the Anderson model.
In the opposite limit, where even a small offset gives a large dipole, we have $\beta \gg 1$, giving $\alpha \approx 1$. Therefore vacuum dipole step of heterostructure is equal and opposite to the difference in midgap energies of constituent semiconductors. Then the final lineup has the midgaps nearly aligned. This is the idea behind the midgap or neutrality-level models proposed for bulk semiconductor heterojunctions by Tejedor, Flores \cite{tejedor1978simple} and Tersoff \cite{tersoff1984theory}. 

\begin{figure}[t]
\includegraphics[width=\columnwidth]{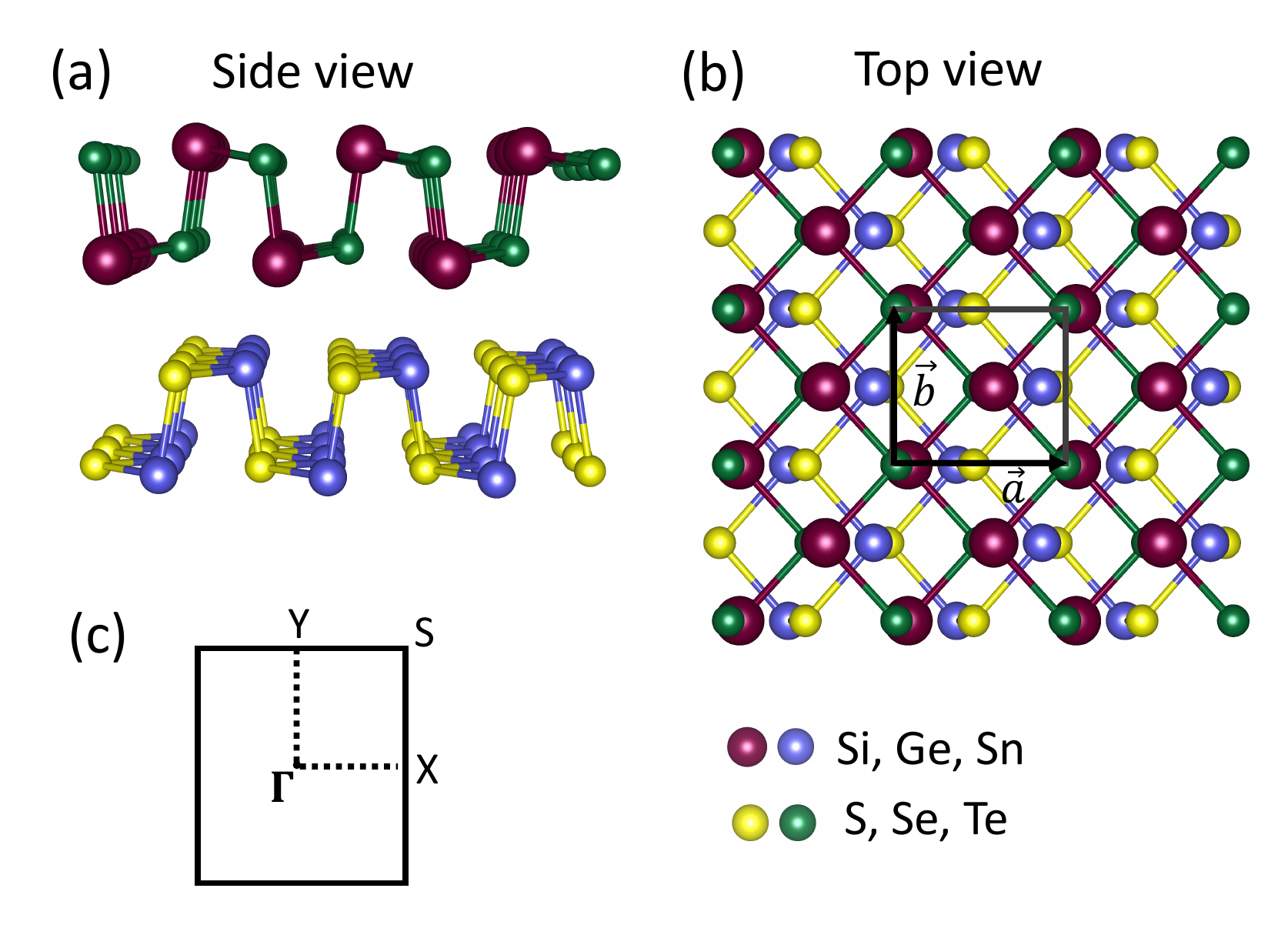}
\caption{\textbf{Group-IV monochalcogenides}. (a) Side view and (b) top view of the optimized atomic structure of group-IV monochalcogenide heterostructure. The rectangular unit cell with in-plane lattice vectors, $\vec{a} $ and $\vec{b} $ are indicated. (c) Corresponding Brillouin zone with symmetry directions.
}
\label{FIG1}
\end{figure}

\section{Density Functional Calculations}

\subsection{Atomic structure}
 Group-IV monochalcogenides have chemical formula of $MX$, where $M$ and $X$ are elements from group-IVA and VIA of the periodic table, respectively. They belong to the space group \textit{Pnma} with an orthorhombic structure in their bulk form. They have a puckered crystal structure similar to the more commonly studied black phosphorous \cite{avouris20172d,li2014black,liu2014phosphorene}. However, unlike black phosphorus, inversion symmetry in the perpendicular direction of the layers is lost for the monolayer, as shown in phosphorene oxides \cite{lee2019low,kang2019two}, and they belong to the space group P\textit{mn}2$_{1}$ instead.     
The primitive unit cell is rectangular with four basis atoms, two from each atomic species. Each atom is covalently bonded to three neighbors of the other atomic species forming zigzag configuration of alternating atoms, and the bonds are typically softer along the armchair direction \cite{ozccelik2018tin}. The atomic configurations of group-IV monochalcogenides vertical heterostructures are illustrated in Fig. 1(a,b). The primitive unit cell of the heterostructure is still rectangular with eight basis atoms and its rectangular Brillouin zone (BZ) with the high symmetry points are shown in Fig. 1(c). Puckered structure of group IV-monochalcogenides is the most stable phase, with formation energies comparable to single-layer MoS$_{2}$ \cite{singh2014computational}.

 \subsection{Computational Method}
We carried out first-principles DFT calculations as implemented in the Vienna \textit{ab initio} simulation package \textsc{(vasp)} \cite{kresse1996efficient}. The Perdew-Burke-Ernzerhof (PBE) functional was chosen within the generalized gradient approximation (GGA) to treat the exchange-correlation interaction of electrons \cite{perdew1996generalized}. The electron-core interaction is described by projector augmented wave (PAW) potentials \cite{blochl1994projector}, and van der Waals corrections have been included in the calculations. The cutoff for plane-wave expansion is set to be 300\,eV. A set of (21$\times$21$\times$1) $k$-point sampling is used for BZ integration in $k$-space, following the scheme proposed by Monkhorst-Pack \cite{monkhorst1976special}. Atomic positions were optimized using the conjugate gradient method, where the total energy and atomic forces were minimized. The energy convergence value between two consecutive steps of $10^{-6}$\,eV was chosen and a maximum of Hellmann-Feynman force of 1\,meV/\AA~ was allowed on each atom. Vacuum spacing of at least 30\,\AA~ are added along the direction perpendicular to the 2D atomic plane for the monolayers and heterostructure, respectively, in order to avoid the interaction between adjacent supercells. In addition to PBE calculations which usually underestimates the bandgaps of semiconductors, for Sec. VI we carried out calculations using HSE06 hybrid functional \cite{paier2006screened} for the free-standing monolayers using the optimized structures obtained by PBE.

\subsection{Monolayers to heterostructure}

We first performed structure optimization of the 9 free-standing $MX$ ($M$ = Si, Ge, Sn; $X$ = S, Se, Te) monolayers and then calculated their electronic band structures. We found that all monolayers are indirect gap semiconductors with bandgaps ranging from 0.34\,eV to 1.55\,eV for PBE and 0.60\,eV to 2.18\,eV for HSE06 calculations. The band edges values for PBE and HSE06 calculations are presented in the Supplemental Material.

The stacking and commensuration of the 2D heterostructure, however, requires more consideration as it depends on the way it was prepared. For example, mechanical exfoliation and transfer methods would typically produce incommensurate bilayer heterostructures with minimal strain \cite{akinwande2014two}. Chemical vapor deposition or molecular-beam epitaxy growth, on the other hand, would admits heterostructures that are commensurate, where the stacking configuration and lattice constants are dictated by the growth conditions and substrates \cite{tongay2014tuning,wang2015all,miwa2015van,barton2015transition}. Strains incurred would certainly have an effect on the band alignment of the heterostructure\cite{arent1989strain}. 

\begin{figure}
\centering
\includegraphics[width=0.38\textwidth]{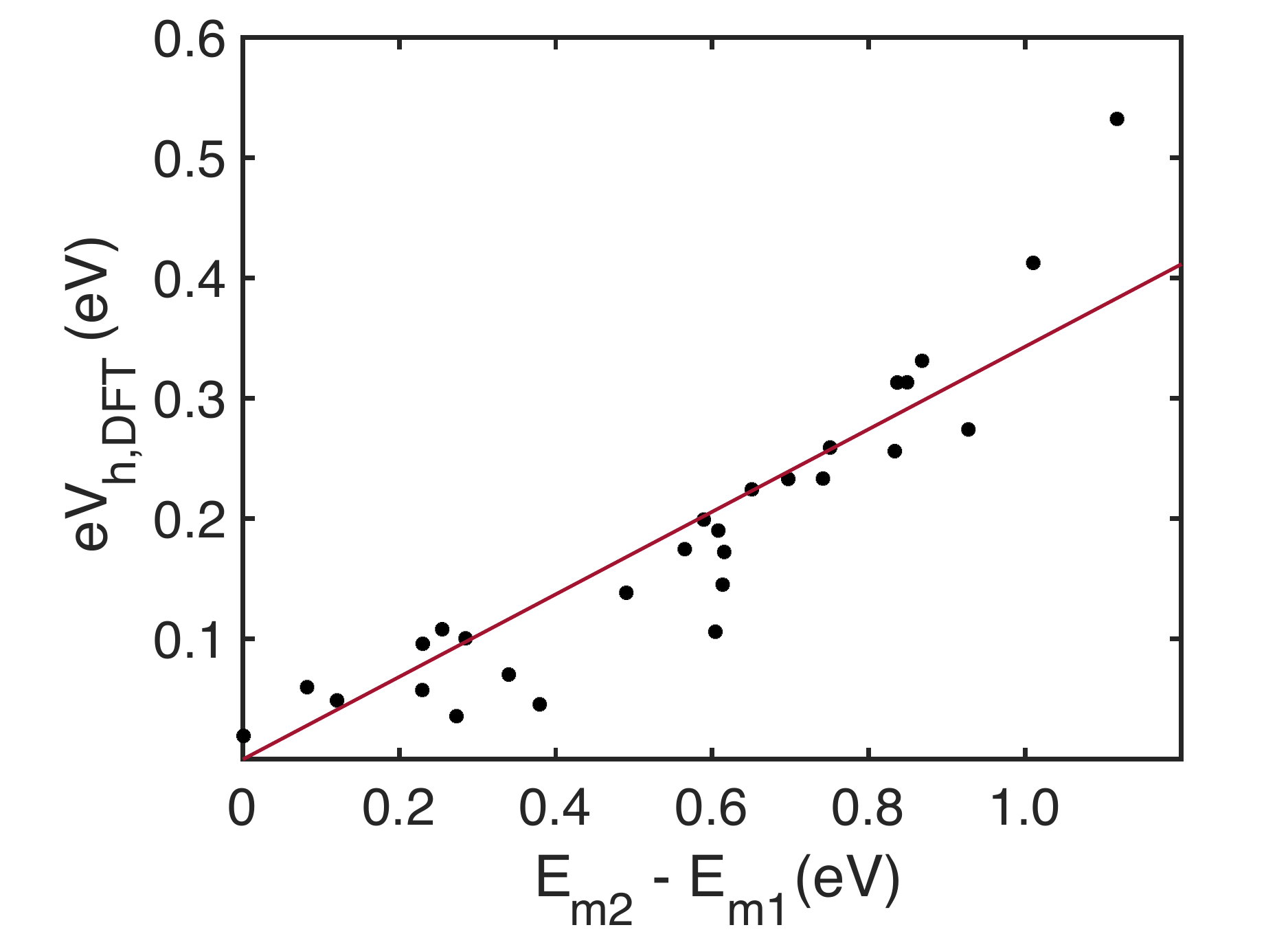}
\caption{\textbf{Extracting linear response coefficient \boldmath$\alpha$}. DFT vacuum dipole step of heterostructures versus midgap energy difference between monolayers (as within the Anderson picture). Linear fitted line with a slope $\alpha=0.34$ is indicated.}
\label{FIG4}
\end{figure}

To extract the linear response coefficient $\alpha$, and to isolate effects of layer interactions, we need to compare monolayers and heterostructures with the same lattice constants. Therefore for each pair $MX$ and $M'X'$, we choose the in-plane lattice constants \textit{a} and \textit{b} to be the larger of the respective lattice constants of the relaxed free-standing monolayers. 
Because SiS monolayer has significantly smaller lattice constants compare to the other $MX$ monolayers [See SI], we do not include heterostructures with SiS in this work.

\begin{figure*}[t!]
\includegraphics[width=18cm]{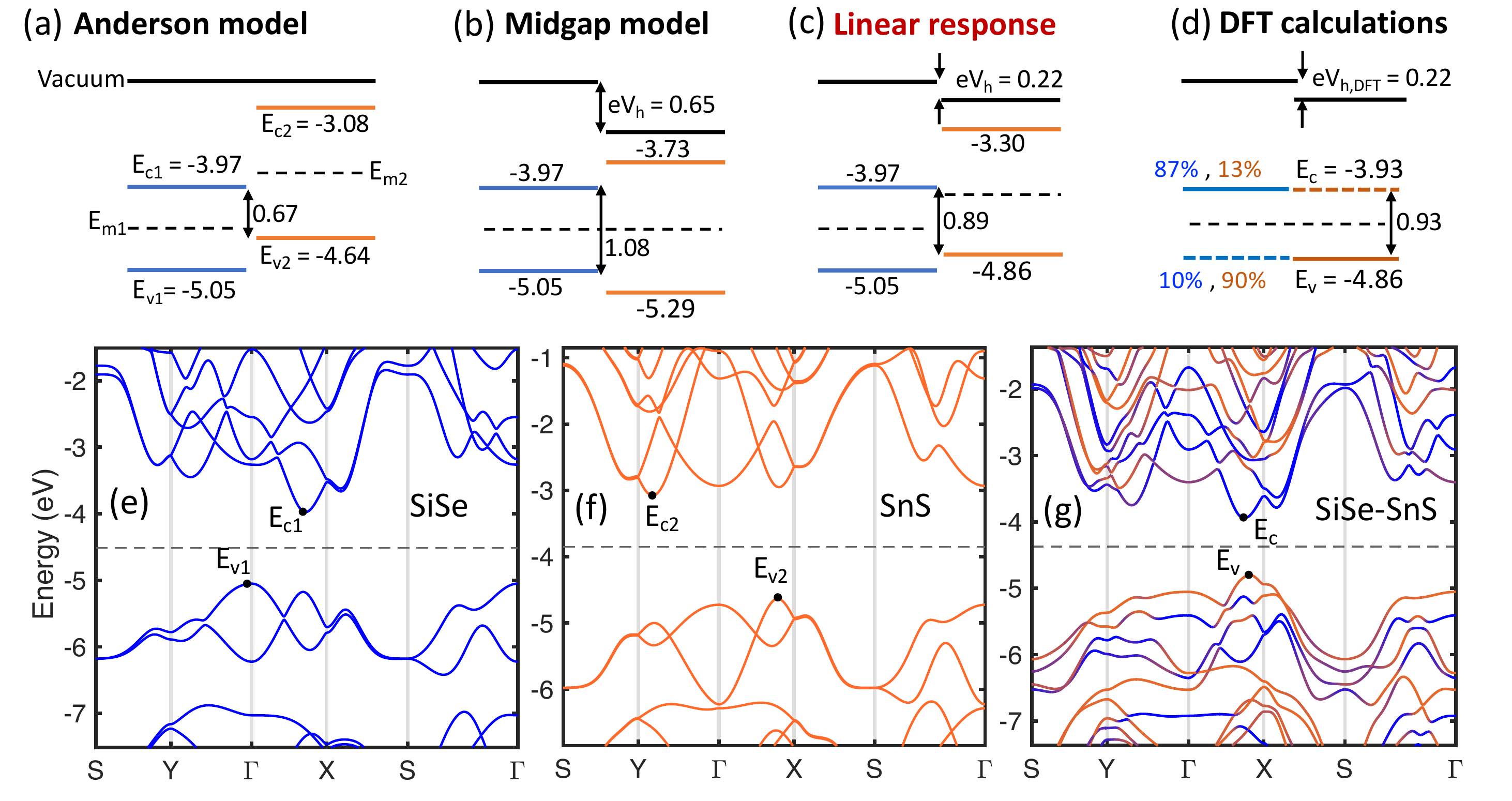}
\caption{\textbf{Comparing energy band models}. (a) Energy band alignment for two isolated 2D monolayers. According to the Anderson model, the heterostructure is a type II with a bandgap of 0.67\,eV. The values for CBM, VBM, and midgap for the first monolayer, SiSe (blue color) and the second monolayer, SnS (orange color) are presented. All energies are in eV.
(b) Energy band alignment in the midgap model, where midgap energies are aligned and energy bands of the second monolayer are shifted to the lower energies, creates a vacuum dipole step of $eV_{h}$. (c) Band model in the linear response model, where vacuum dipole is obtained by $eV_{h} = 0.34\times(E_{m2}-E_{m1})$ = 0.22\,eV. This leads to a type II heterostructure with a bandgap of 0.89\,eV, in a great agreement with the DFT. (d) Energy band diagram for the DFT calculated SiSe-SnS heterostructure with a vacuum dipole step of $eV_{h,\mathrm{DFT}}$. DFT bandgap is 0.93\,eV. The percentages represent the contribution of each layer to the band edges. 
DFT band structure of isolated (e) SiSe, (f) SnS monolayers, and (g) SiSe-SnS heterostructure. CBM and VBM values are indicated by dots. The midgap energies are illustrated by horizontal dashed lines. All energies are taken with respect to the vacuum level (For the heterostructure, we used the highest vacuum level).
} 
\label{FIG2}
\end{figure*}

\subsection{Extracting linear response coefficient}
In general, $\alpha$ may be different for every bilayer. To give a predictive model, we make the somewhat drastic approximation that $\alpha$ is the same for all bilayers considered here. Then we determine $\alpha$ by fitting to the bilayer dipoles calculated in DFT. Figure 2 shows the vacuum dipole step $eV_{h,\mathrm{DFT}}$ (reflecting the sheet dipole moment) of each heterostructure, versus $E_{m2} - E_{m1}$, the difference in midgap energy (relative to vacuum) of the respective monolayers. A typical vacuum step, and how that is extracted from DFT, is shown in Fig. S1 of the SI. The layers are numbered so that $E_{m2} - E_{m1}$ and thus $eV_{h,\mathrm{DFT}}$ have positive values. The best linear fit for this set of materials corresponds to $\alpha=0.34$. This suggests that the actual band alignment is intermediate between the Anderson and midgap models. We examine the accuracy and consequences in more detail in the subsequent sections. We should note that for each family of 2D materials, a different $\alpha$ is expected, and can be determined using the same procedure.

\section{Band alignment models}

For bulk semiconductor heterostructures, the band lineups are classified into three types \cite{ozcelik2016band}. In type I (straddling) heterostructure, both conduction band minimum (CBM) and valence band maximum (VBM) resides in the same semiconductor. For type II (staggered), the CBM and VBM of the one semiconductor have higher energies than those of the other. In type III (broken-gap), the CBM of one semiconductor energetically falls below the VBM of the other semiconductor. Since vdW heterostructures can have very weak hybridization between the layers, we adopt the same classification here, based on the monolayer bands shifted by the vacuum dipole. This gives a model band alignment, valid in the limit of weak hybridization between layers. In this way we can compare various models to the DFT results. 

\begin{figure*}[t!]
\includegraphics[width=18cm]{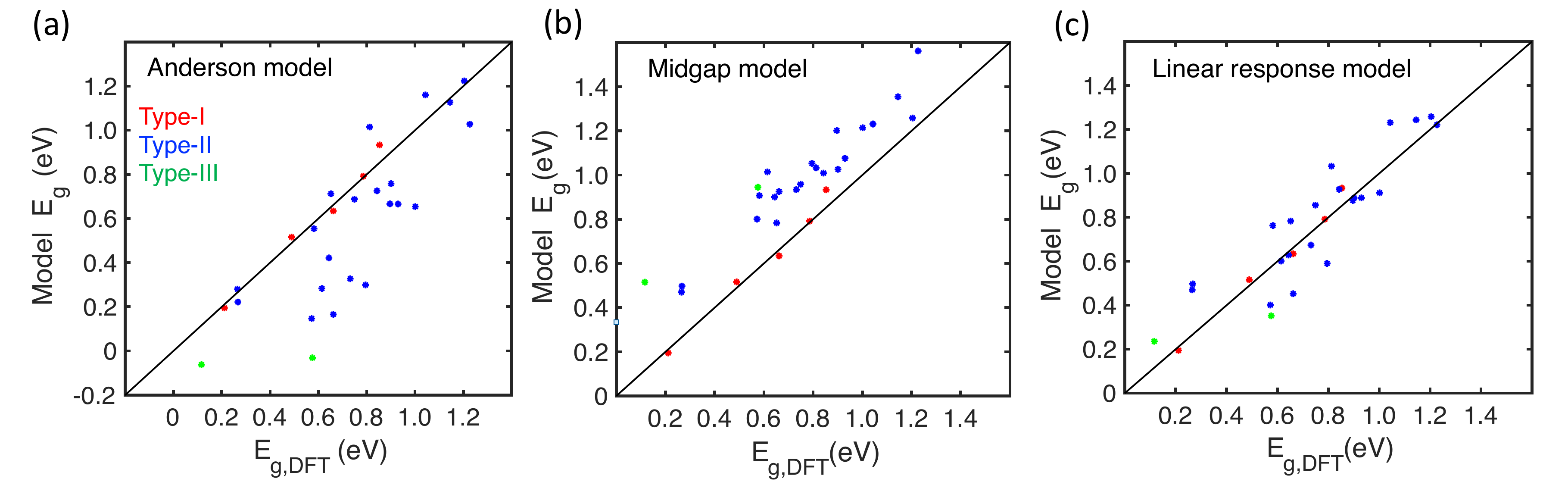}
\caption{ \textbf{Survey of bandgaps across the models}. (a) Comparison between the bandgaps extracted from the Anderson model, and that computed from DFT, $E_{g,\mathrm{DFT}}$. (b) and (c) same as (a) but for the midgap and linear response model, respectively. }
\label{FIG4}
\end{figure*}

Figure 3(a) shows a heterostructure formed by SiSe and SnS monolayers. SiSe (SnS) monolayer band edges are represented with blue (orange) lines in Fig. 3(a), with its CBM $E_{c1}$ ($E_{c2}$), and VBM $E_{v1}$ ($E_{v2}$), all with respect to the vacuum level. DFT band structure of monolayers and their heterostructure are presented in Fig. 3(e-g), where CBM and VBM are indicated in the plots. The midgap energies of SiSe and SnS are denoted as $E_{m1}$ and $E_{m2}$, shown by horizontal dashed lines. Notice that in this paper we use $E_{g}$ for the predicted bandgaps of the heterostructures based on the models, and $E_{g,\mathrm{DFT}}$ for the DFT calculated bandgaps for the heterostructures. \\

The predicted bandgap of the heterostructure, $E_{g}$, would be given by the difference between the lowest lying conduction band and the highest lying valence band. For the SiSe-SnS heterostructure, bandgap according to the Anderson model is $E_{g}$ = 0.67\,eV, as shown in Fig. 3(a).

Figure 3(b) depicts the band alignment according to the midgap model. 
Predicted bandgap based on the midgap model, $E_{g}$, is simply given by the smaller bandgaps of the two monolayers, which in this case would be $E_{g}$ = 1.08\,eV. This overestimates the DFT calculated bandgap of $E_{g,\mathrm{DFT}}$= 0.93\,eV, as shown in Fig. 3(d).

Figure 3(c) shows the energy band alignment in the linear response model. As discussed in Sec. III D, linear response coefficient for group-IV monoahclcogenides is found to be 0.34, which results in a vacuum step of $eV_{h} = 0.34\times(-3.86+4.51) = 0.22$\,eV for the SiSe-SnS heterostructure. This lowering of the bands of SnS monolayer gives a type II heterostructure with $E_{g}$ = 0.89\,eV.
To determine the DFT band alignment, we assign each band edge to SiSe or SnS according to its wave functions. As shown in Fig. 3(d), CBM (VBM) is formed by 87\% (10\%) SiSe contributions. Therefore, we assign the CBM and VBM of the heterostructure to the SiSe and SnS monolayers, respectively. This allows us to assign a type II band alignment in DFT. Thus in this instance, the dipole, bandgap and band type in the linear response model agree well with the DFT calculations.

\section{SURVEY OF MONOCHALCOGENIDES HETEROSTRUCTURES}
\subsection{Bandgaps}

Following the procedures in Sec. IV, we obtained the predicted bandgaps of heterostructures for the group-IV monochalcogenides heterostructures according to the Anderson, midgap, and linear response models, and compared them against the DFT bandgaps.
The results are presented in Fig. 4, where each heterostructure is color coded according to its band alignment in the Anderson picture.
\begin{figure*}[t!]
\centering
\includegraphics[width=\linewidth]{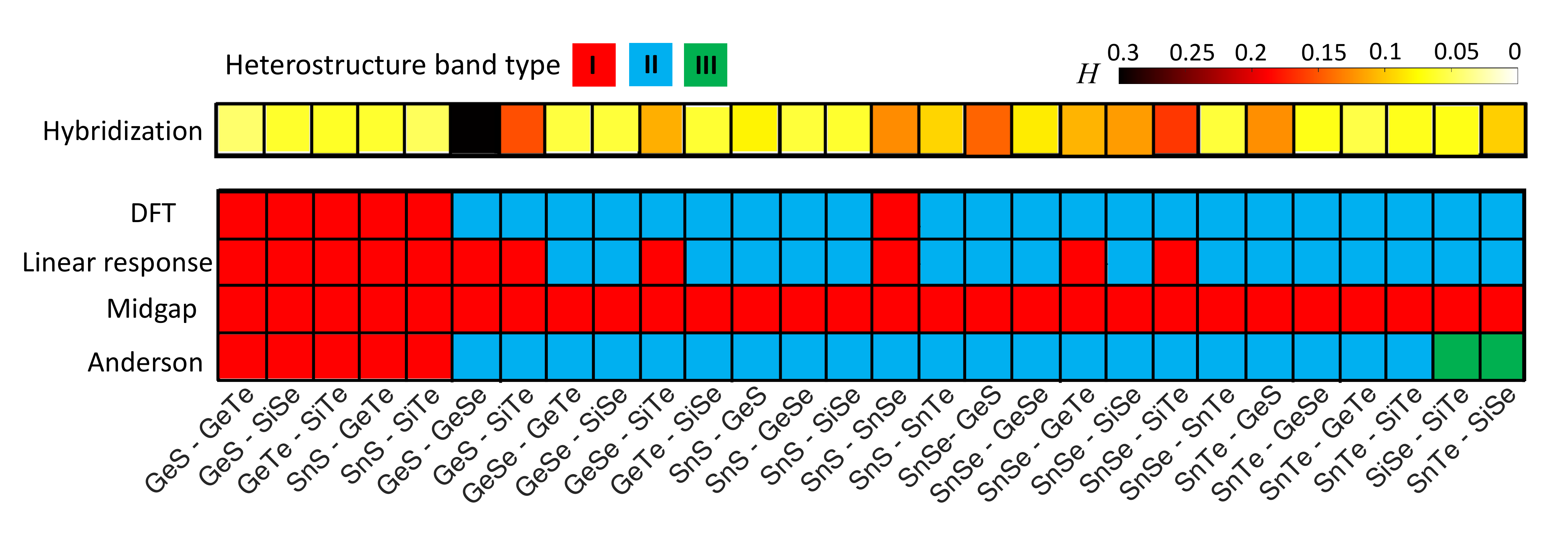}
\caption{\textbf{Survey of the band alignment types}. Comparison of band types for IV-VI heterostructures according to the Anderson, midgap, linear response models, and DFT calculations. Band hybridization strength (\textit{H}) displayed as a heatmap (top row).}
\label{FIG5}
\end{figure*}
Heterostructures which are type I in the Anderson picture are automatically type I in all three models, and these show good agreement with the corresponding DFT values as indicated by the red dots. We note that the band hybridization at the band edges in the type I heterostructures studied in this work is weak, which could be another indication of the great agreement between the predicted and DFT bandgaps.

Figure 4(a) shows that the Anderson model systematically underestimates the bandgaps for the majority of the heterostructures. For type III alignment, there is no bandgap, and the degree of band overlap is report as a negative bandgap in Fig 4.
The midgap model, in contrast, systematically overestimates the bilayer bandgaps. The linear response model gives a balanced error, with better overall accuracy.
To quantify the accuracy, we calculated mean absolute error (MAE) between the model predictions and DFT calculated bandgaps, $\frac{1}{n} \sum_{i=1}^{n}\lvert E_{g}-E_{g,\mathrm{DFT}}\rvert$.

We obtained MAE of 0.18\,eV and 0.20\,eV for the Anderson and midgap models, while the linear response model MAE is only 0.10\,eV. The results show that our proposed linear response model is the most accurate in predicting the bandgaps for group-IV monochalcogenides heterostructures. This is not surprising, since it incorporates additional experimental information via $\alpha$. But it is striking that accuracy is improved by a factor of 2 using only a single $\alpha$ value to describe systems with a wide range of bandgaps, spanning 3 rows of the periodic table.

\subsection{Band alignments}

We complete our survey of monochalcogenides heterostructures by comparing heterostructure band types obtained from different models against DFT results, as shown in Fig. 5. We determine the types of band alignment based on the DFT results by comparing the wave function projections to each layer at the CBM and VBM of heterostructures (see Fig. 3(d)). As shown in Fig. 5, the two heterostructures that had type III band alignment in the Anderson model were found to be either type I or II in the linear response model and DFT calculations. Both the linear response model and DFT do not show any type III, which by definition will be a semimetal without a bandgap. As mentioned before, the midgap model always predicts type I heterostructures. 
In Fig. 5 we see that the linear response model gives the same band type as DFT, except for five heterostructures. 

Strictly speaking, band offsets and band types are concepts appropriate for interfaces between 3D semiconductors. These concepts are useful for 2D systems only insofar as the hybridization is weak. In order to quantify band hybridization in our 2D heterostructures, we computed wave function projections to each layer of the heterostructures for all relevant energy bands.
From the wave function $\Psi_{n,\textit{k}}(\vec{r})$ of band $n$ and $\textit{k}$ point corresponding to the CBM or VBM, we obtained the projection of the wave function to the first layer as 

\begin{equation}
\phi_{1} = \int_{z_{\mathrm{min}}}^{z_{0}}dz\int_{2D}^{}dxdy \lvert\Psi_{n,\textit{k}}(\vec{r})\rvert^{2} 
\end{equation}
 where $z_{\mathrm{min}}$ is the lower boundary of the supercell in the out-of-plane direction and $z_{0}$ is the midpoint position between two layers.
 For the second layer, $z_{0}$ and $z_{\mathrm{max}}$ are the lower and upper integral limits, respectively.
 Now, we define a figure-of-merit for band hybridization, $H$ as
 
 \begin{equation}
H =  |\phi_{1,c}-1/2|^{-1} + |\phi_{1,v}-1/2|^{-1}
\end{equation}
where $\phi_{1,c}$ and $\phi_{1,v}$ are wave function projections of the first monolayer to the CBM and VBM, respectively.
A heatmap of $H$ values for all heterostructures is summarized in Fig. 5, where large $H$ indicates strong band hybridization between layers.
We see that the linear response model is most likely to predict incorrect band type for heterostructures with strong band hybridization.
This is reasonable, since strong hybridization makes the assignment of band type less meaningful even within DFT. In addition, since band hybridization results in the formation of new bonding and antibonding states, it is reasonable to expect that for cases with strong hybridization, the predicted bandgaps differ from the DFT bandgaps. Indeed, our results supports this conjecture.

For example, GeS-SiTe and SnSe-GeTe are heterostructures with especially strong band hybridization. For these, our linear response model fails to predict the correct band type, and they also exhibit the largest difference between the predicted and DFT bandgaps.

The results of this section show that the midgap model is incapable of predicting band type, and the Anderson model underestimates the bandgap significantly, but the linear response model correctly predicts both bandgap and band type of the selected heterostructure, and it is indeed an improvement over the other two models as compared to DFT.

\section{Application of linear response model}

\begin{figure}[t!]
\includegraphics[width=\columnwidth]{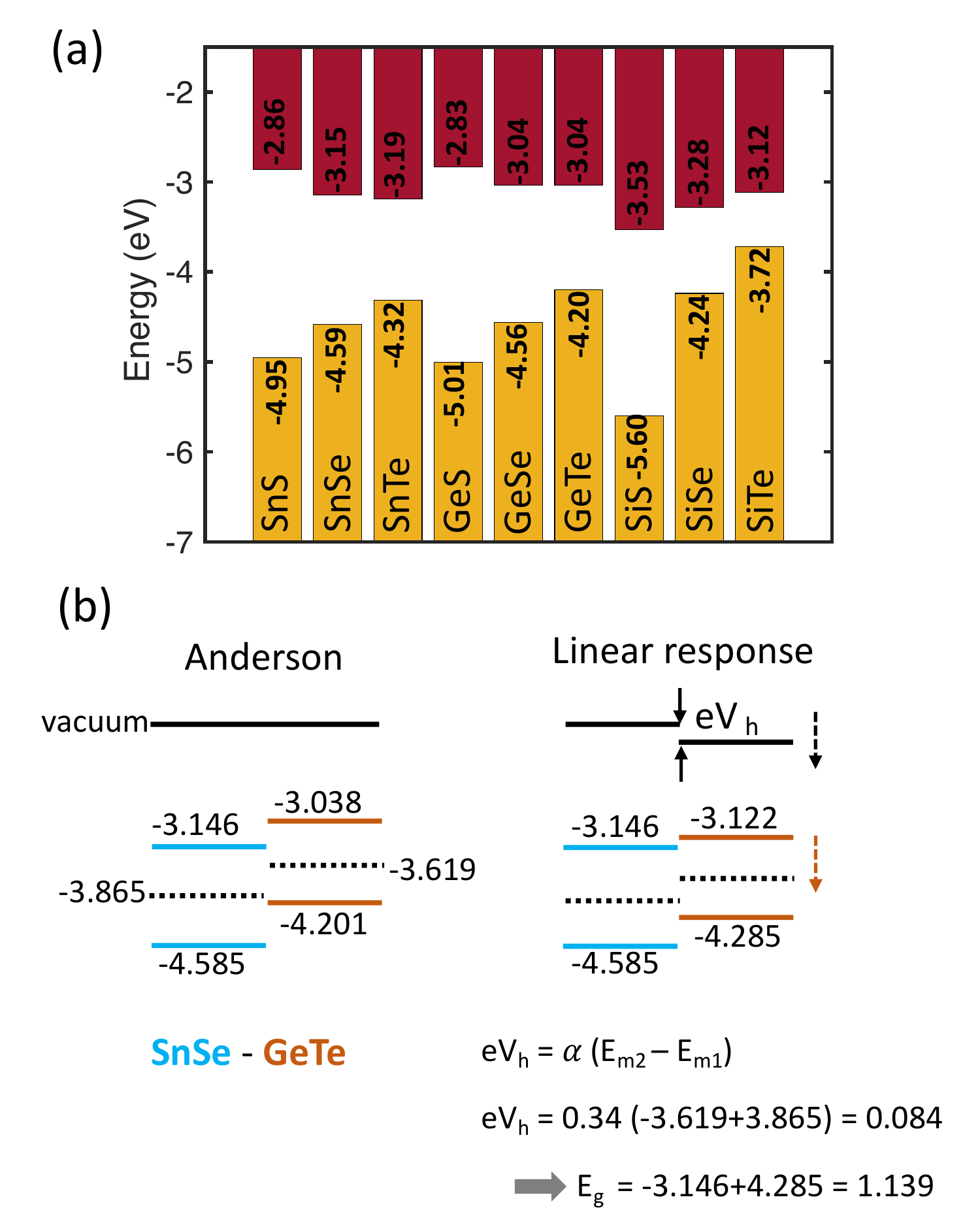}
\caption{\textbf{Application of the linear response model}. (a) HSE06 calculated band edges of free-standing group-IV monochalcogenides monolayers, relative to the vacuum energy. CBM and VBM values are indicated. An example; (b) Band lineup of SnSe-GeTe in the Anderson and linear response models. SnSe and GeTe band edges are illustrated by blue and orange colors, respectively. Vacuum dipole step of 0.084\,eV is computed from linear response model as illustrated, which yields a bandgap of 1.139\,eV for the heterostructure. All energies are in eV.}
\label{FIG5}
\end{figure}

The ultimate goal is not only to understand the role of dipole formation, but to predict experimental lineups.  For this purpose, we should calculate the monolayer properties in whatever state of strain occurs in the experiment, and then shift them by the dipole. A common case, and the simplest case, is that both layers are unstrained. This occurs when individual layers are mechanically exfoliated and combined. For predicting experimental values, we would like to have more accurate band gaps than are provided by DFT, which systematically underestimates bandgaps \cite{wang1983density}. Hybrid density functionals have been shown to improve the description of bandgap of semiconductors \cite{muscat2001prediction,chan2010efficient,perdew2017understanding,tran2017importance}, so we use the HSE06 hybrid functional \cite{chan2010efficient} here. Once we calculate the unstrained monolayer properties within HSE06, we use those within the linear response model to calculate the band lineup. We neglect any change in the response coefficient associated with changing the strain state or going from DFT to HSE06. This is reasonable, since in any case it is an approximation to use a single value of $\alpha$ for the whole family of materials treated here.

The calculated band edges of the relaxed monolayers relative to their vacuum energies are shown in Fig. 6(a), where maroon and yellow bars show conduction and valence bands, respectively.
From $eV_{h} = \alpha \times(E_{m2}-E_{m1})$, the vacuum dipole steps of the heterostructures can be computed for any pair of these monolayers, using the value $\alpha=0.34$ determined above. The bandgap of the heterostructure is given by $E_{g} =  \mbox{min}(E_{c1}, E_{c2} - eV_{h})- \mbox{max}(E_{v1}, E_{v2} - eV_{h})$, where second layer is always chosen to have higher midgap energy ($E_{m2}$) than the first one ($E_{m1}$).
We take SnSe and GeTe monolayers as an example. The midgap energies of SnSe and GeTe monolayers are $E_{m1}=-3.865$ and $E_{m2}=-3.619$\,eV, respectively [Fig. 6(b)]. Therefore, for this case $eV_{h} = 0.34 \times(-3.619+3.865) = 0.084$\,eV. Then, the bandgap of SnSe-GeTe heterostructure according to the linear response model is given by $E_{g} = 1.139$\,eV as illustrated. These two monolayers form a type II heterostructure, as shown in Fig. 6(b). 
Using the same approach, we obtained bandgaps of all the heterostructures. These results are listed in Table I of SI.

Using the linear response model, the bandgaps range from 0.17\,eV for SiSe-SiTe to 1.51\,eV for SnS-GeS heterostructure. We found that only SnTe-GeTe, SnSe-GeTe, and SiSe-SiTe are type II, while the other heterostructures are type I. These results can be compared against experimental data based on mechanically stacked vdW heterostructures, when those become available.

\section{Conclusion}
A simple linear response model allows us to predict the electronic bandgap of 2D vertical heterostructures using only the constituent monolayers band edges. Tests using DFT and idealize strain states demonstrate that the model provides significant improvement relative to the popular Anderson and midgap models. We also found that the model is most accurate when band hybridization between the two monolayers is small. In addition, by using more accurate HSE06 calculations for free-standing monolayers, we predict the bandgaps and band types of the vertically-stacked group IV-VI unstrained heterostructures. The linear response model can be applied to other families of 2D materials, and thus enable experimentalists and materials scientists to screen and select favorable materials with the targeted band types and bandgaps for desired applications.

\section*{Acknowledgements}
JA and TL acknowledge partial funding support from NSF DMREF-1921629 and in part by NSF ECCS-1542202. SL, HRK, and YKK acknowledge financial support from the Korean government through the National Research Foundation (NRF) of Korea (No. 2019R1A2C1005417). We acknowledge computational support from the Minnesota Supercomputing Institute (MSI). Some portion of our computational work was done using the resources of the KISTI Supercomputing Center
(KSC-2020-CRE-0011).

\bibliographystyle{unsrt}

\end{document}